# Generative model for information metamaterial design

Jun Ming Hou[1,2,6], Long Chen[1,2,6], Xuan Zheng[1,2,6], Jia Wei Wu[1,2,6], Jian Wei You[1,2,*], Zi Xuan Cai[1,2], Jiahan Huang[1,2], Chen Xu Wu[1,2], Jian Lin Su[1,2], Lianlin Li[3], Jia Nan Zhang[1,2,*], and Tie Jun Cui[1,2,4,5,*]

[1] State Key Laboratory of Millimeter Waves, Southeast University, Nanjing 211189, China
[2] School of Information Science and Engineering, Southeast University, Nanjing 211189, China
[3] State Key Laboratory of Photonics and Communications, Peking University, Beijing 100871, China
[4] Institute of Electromagnetic Space, Southeast University, Nanjing 211189, China
[5] Suzhou Laboratory, Suzhou 215164, China
[6] These authors contributed equally to this work.
[*] Email: jvyou@seu.edu.cn, jiananzhang@seu.edu.cn, and tjcui@seu.edu.cn

## Abstract

Generative models such as AlphaFold[1-2] and MatterGen[3] can directly generate novel material structures with desired properties, accelerating the new materials discovery and revolutionizing the material design paradigm from traditional trial-and-error approach to intelligent on-demand generation. AlphaFold is focused on protein prediction with specific aperiodic structures; while MatterGen is focused on predicting periodic and stable crystal structures. The universal design of metamaterials is much more complicated[4-11], since it involves to design meta-atoms (similar to the periodic structures) and their arbitrarily inhomogeneous distributions in space. Here, we propose InfoMetaGen, a universal generative model for information metamaterial design, which combines a pre-trained foundation model with lightweight functional adapters to intelligently generate artificial structures on-demand spanning from meta-atoms to arbitrary space coding patterns. In contrast to conventional intelligent metamaterial design methods that require training dedicated models for specific functionalities, InfoMetaGen enables a single universal generative model capable of switching across diverse functionalities by fine-tuning the lightweight adapters, significantly improving both efficiency and generalizability. Experimental results demonstrate that InfoMetaGen can not only accelerate the diverse discovery of new metamaterials, but also achieve breakthroughs in metamaterial performance. This work fills the gap of universal generative framework in designing artificial materials, and opens up unprecedented opportunities to expand the capability of generative models from the passive discovery of microscopic natural material to the active creation of macroscopic artificial materials.

# Introduction

Generative artificial intelligence (AI) models are revolutionizing the fundamental paradigm of scientific discovery, demonstrating unprecedented potentials in structural biology and materials design. For instance, AlphaFold[1,2] has achieved groundbreaking success in aperiodic structure of protein prediction, providing a powerful tool for medicine design and life sciences research; and MatterGen[3] employed a diffusion model to directly generate novel, periodic and stable crystal structures targeting specifically desired material properties. The emergence of these generative models marks a paradigm shift in material discovery from traditional "trial-and-error" screening to a "generative" design, significantly accelerating new material discovery. These generative models greatly enlarge the material exploration space that is far beyond the limited set of known materials, and enable us to explore the previously inaccessible regions. However, AlphaFold is focused on protein predictions with specific aperiodic structures; while MatterGen is focused on predicting periodic and stable crystal structures. The universal design of materials requires to generate arbitrarily distributed molecules, composites or structures, which is a bottleneck for the current generative models.

Metamaterials are a class of macroscopic artificial materials whose physical properties are not determined by their chemical composition but are governed by precisely designed meta-atoms and their spatial distributions[4-11]. These artificial structures enable the realization of exotic physical properties absent in natural materials, such as negative refractive index and negative Poisson's ratio. In addition to discover novel and counter-intuitive physical phenomena, the metamaterials have found widespread applications in wireless communications, wireless sensing, and super-resolution imaging, drawing considerable attention from both physics and engineering communities. In order to control electromagnetic fields and waves in real time and improve the functional reconfigurability[12-17] of metamaterials, the concept of digital codingmetamaterials was proposed[18], creating field-programmable metamaterials[19-23]. Subsequently, they are evolved into information metamaterials[24-30], bridging the electromagnetic physics to digital information and enabling the transformation of metamaterials from effective materials into advanced information processing platforms.

The success of metamaterials lies in their structure-determines-function features, which enable the precise controls over the responses of materials to the excitation of physical fields by artificially designing their structural configurations. This characteristic allows metamaterials to achieve some properties and functionalities that are difficult or impossible to attain with natural materials. However, the structure-determines-function feature introduces significant complexity challenges to design metamaterials. Due to the extremely vast combinatorial space of structural parameters, the traditional design approaches heavily rely on empirical knowledge and trial-and-error iterations, resulting in prolonged design cycles, limited flexibility and inadequate efficiency. To address these challenges, intelligent methods are emerging as a new paradigm in designing metamaterials[31-38]. By introducing machine-learning design models, it is possible to substantially shorten the design cycles, enable efficient optimizations, and facilitate diverse generations of the metamaterial structural configurations. Nevertheless, the intelligent designs of metamaterials remain in a relatively nascent stage, and a universal framework of generative design suitable for the metamaterials has not yet been established, since it involves to design the meta-atoms and their arbitrarily inhomogeneous distributions in the space.

In this study, we propose a universal generative model to design information metamaterials, InfoMetaGen, which can generate multi-bit meta-atoms and arbitrarily inhomogeneous coding sequences for meta-array with required functions. InfoMetaGen demonstrates strong adaptability to a wide range of downstream tasks for the inverse design of metamaterials. In designing the meta-atoms, we incorporate a diffusion-based process that parameterizes the meta-atoms based on structural features such as periodicity and symmetry, combined with adapter modules to tune the desired electromagnetic responses precisely, including the frequency, phase and amplitude. Compared to the prior approaches, InfoMetaGen demonstrates superior generative capacity by rapidly producing a large number of diverse coding patterns that satisfy the specific functional requirements. It exhibits strong extrapolation ability, generating novel meta-atoms beyond the training dataset and identifying high-performance configurations. In designing the meta-array with inhomogenous space coding sequences, the high degree of controllability of InfoMetaGen greatly enhances the functional flexibility by enabling precise customization without requiring the task-specific model training. Instead, it enables a multi-task design through fine-tuning, and support the functionalities such as beam steering, electromagnetic focusing, and holographic imaging. It is worth noting that our model can generate 1-bit and 3-bit inhomogeneous meta-arrays as needed, which has not been achieved in previous studies. To verify the effectiveness of InfoMetaGen, we conduct experimental evaluations in both meta-atom and meta-array levels. The experimental results strongly confirm the universal generation capability of InfoMetaGen and its performance for applicability.

## Generative model for information metamaterial design

The metamaterials are a class of artificial materials that serve as the macroscopic counterpart to microscopic natural materials, with the meta-atoms functioning analogously to atoms and the meta-array corresponding to natural materials, as shown in **Fig. 1a**. Although current generative models have achieved great success in designing the microscopic natural materials, they have not yet been effectively applied to the macroscopic artificial materials. To address this gap, we propose

InfoMetaGen, a diffusion-based generative model for the inverse design of information metamaterials, capable of generating both the structural configurations of individual meta-atoms and the whole meta-arrays.

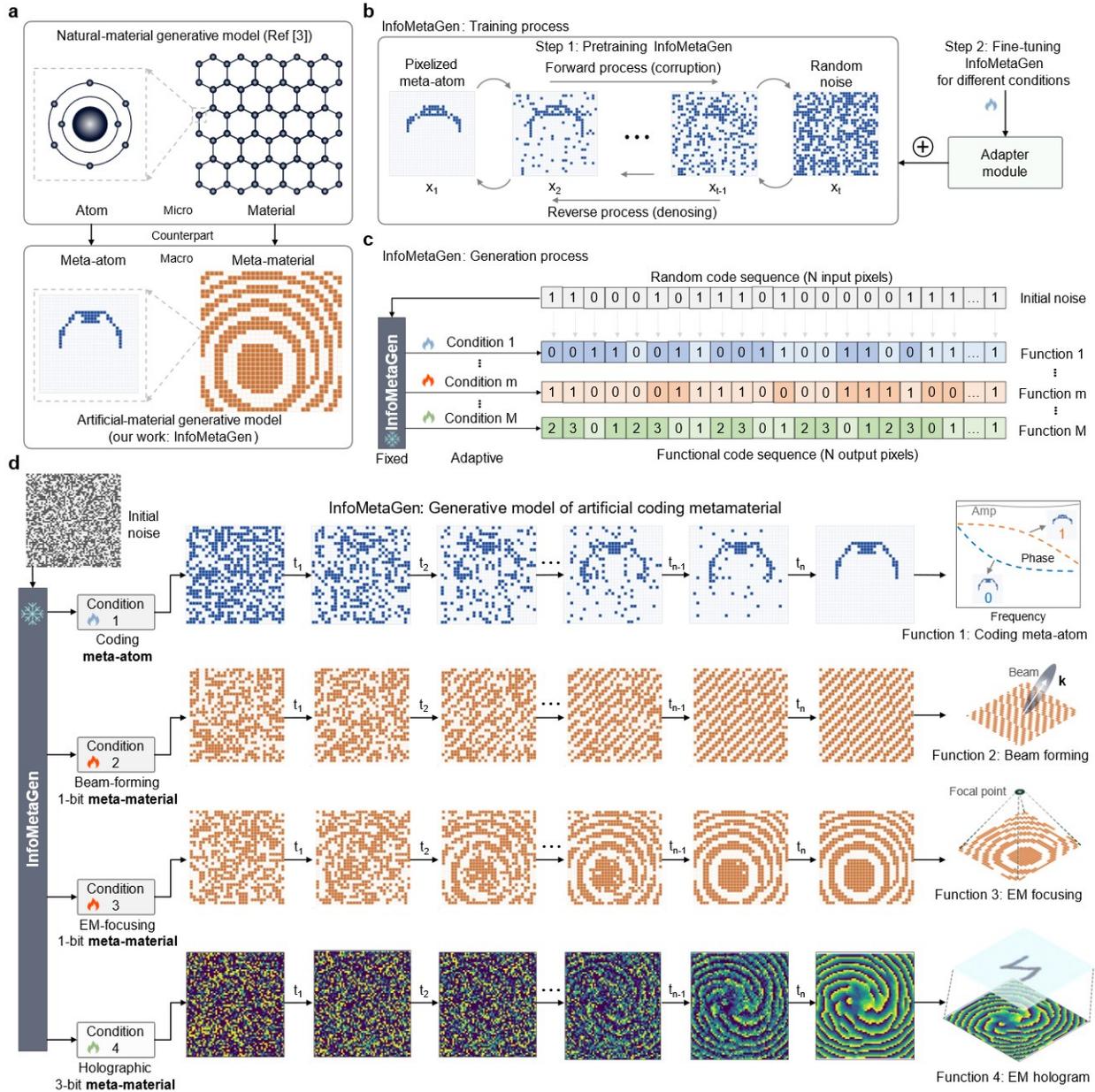

**Fig. 1 | Intelligent design of information metamaterials based on InfoMetaGen. a,** Analogy between the microscopic natural materials and macroscopic artificial metamaterials. **b,** Training process of InfoMetaGen. During the forward diffusion process of the base model, the functional coding pattern gradually degrades into a random noise. The reverse process progressively restores the original functional coding pattern. For different

design tasks, the base model can be adapted through fine-tuning with an adapter module. **c,** Generation process of InfoMetaGen. Based on the same base model, the corresponding coding bitstreams can be flexibly generated by adjusting the conditions according to different functional tasks. **d,** Application of InfoMetaGen. It enables the unified generation of tasks across distinct applications, including the meta-atom design, beam steering, electromagnetic focusing, and holographic imaging.

The proposed InfoMetaGen framework employs a two-stage training strategy. Firstly, an unconditional diffusion model is pretrained to capture the functional coding patterns in diverse design spaces, including different functions and bit-resolutions. Secondly, a conditional adapter module is introduced to finely tune the pretrained diffusion model for guiding the generation process toward task-specific coding patterns under the given conditions. As depicted in **Fig. 1b**, we outline the training process of InfoMetaGen. Given distinctive digital coding characteristics of the information metamaterials, both meta-atoms and meta-arrays can be conceptualized as coding patterns or as discrete digital bit streams. In the forward process, the encoding matrices from different functionalities and bit numbers are randomly sampled and converted into analog bit representations compatible with the continuous diffusion framework, and then progressively perturbed with Gaussian noise to corrupt them toward a random distribution. In the reverse process, the corrupted analog matrices are iteratively denoised and mapped back to the corresponding discrete encoding patterns. In addition, we introduce a functionality-oriented conditional adapter that finely tunes the pretrained diffusion model, enabling inverse design of metamaterial under the desired conditions. Once the base diffusion model is pretrained, we only need to finely tune the adapter modules for the target functionalities. These tunable components are integrated in each layer of the base model, enabling dynamic adjustment of outputs according to the given functional conditions. Importantly, for a specific function, only a single conditional adapter is required to handle different bit resolutions, eliminating the need to train separate adapters for bit setting. This design allows efficient adaptation to diverse functional requirements while keeping the base model parameters fixed, ensuring stability and scalability of the generation process.

By providing specific functional conditions, a set of random noise is rapidly transformed to a functional digital bit stream, as illustrated in **Fig. 1c**. This functionality-oriented fine-tuning design substantially reduces the computational costs while enhancing the model's flexibility and applicability. **Figure 1d** shows several practical applications of the well-pretrained InfoMetaGen base model. By providing specific functional conditions, users can achieve the generation of four distinct functionalities from the noise to digital encoding matrices: 1) design of meta-atoms that meets the amplitude and phase requirements at specific frequencies; 2) beam steering at specific three-dimensional (3D) angles; 3) electromagnetic focusing at any 3D spatial locations, and 4) holographic imaging of arbitrary images. InfoMetaGen not only establishes a universal model for various metamaterial functions but also presents new possibilities for solving the inverse design problem in artificial materials.

**Generating new meta-atoms**

The uniqueness of metamaterials stems from their capability to modulate multiple degrees of freedom of electromagnetic waves independently, which fundamentally relies on the features of meta-atoms, particularly their amplitude and phase responses across different frequencies. Hence we begin the application of InfoMetaGen in designing new meta-atoms, the fundamental units to control the electromagnetic waves of information metamaterials and analogous to the atoms in natural materials.

To demonstrate the capability of InfoMetaGen in generating novel, distinct and symmetric meta-atoms, we discretize the meta-atom into a pixelated structure, as shown in **Fig. 2a**. In this discretized representation, metallic regions are assigned to value of 1, while non-metallic regions are assigned to value of 0. It is crucial to highlight that the binary bit stream of 0 and 1 in meta-atom design fundamentally differs from the 1-bit coding scheme of 0 and 1 used in meta-array encoding. In the context of encoding the meta-array, the states 0 and 1 correspond to phase bits with the phase difference of 180°; that is to say, they represent two meta-atoms that exhibit complementary phase responses with opposite electromagnetic characteristics. Here, in the meta-atom design, we select a transmission-type Huygens meta-atom structure for discretization, and use its phase responses and the target working frequency as conditions to finely tune the adapter. We remark that the design of transmission-type metamaterials is much more difficult than that of reflection-type. Using the finely tuned model, we demonstrate that InfoMetaGen can generate the anti-phase coding meta-atoms working at the target frequency of 17 GHz. As a conceptual validation, we conduct experimental tests, and the good agreement between simulated and measured results of the phase-frequency curves demonstrates our model's ability to generate the desired coding meta-atoms, as shown in **Fig. 2a**, where the inset illustrates the photographs of two measured 1-bit meta-atoms. Considering the broadband characteristics of Huygens' metamaterials, we further validate the model's capability for broadband generations, including the target frequencies at 8 GHz, 9 GHz, 11 GHz and 12 GHz, as illustrated in **Fig. 2b**.

To quantitatively assess the InfoMetaGen's generative capability for meta-atom designs, we define four evaluation criteria: "New" indicates that the performance of the generated meta-atoms is not present in the previous dataset; "Unique" signifies that the structure of the generated meta-atoms is previously absent in the dataset; "OOD" represents that the generated meta-atoms are outside of the existing dataset; and "Symmetric" denotes whether the generated meta-atom structure satisfies the required symmetry constraint. As shown in **Fig. 2c**, these four evaluation metrics remain consistently high across 6,000 generations, demonstrating the model's robust ability to generate new, unique and symmetric meta-atoms. **Figure 2d** shows that InfoMetaGen is capable of generating meta-atoms beyond the scope of the existing database, although this process requires more iterations than for those within the dataset.

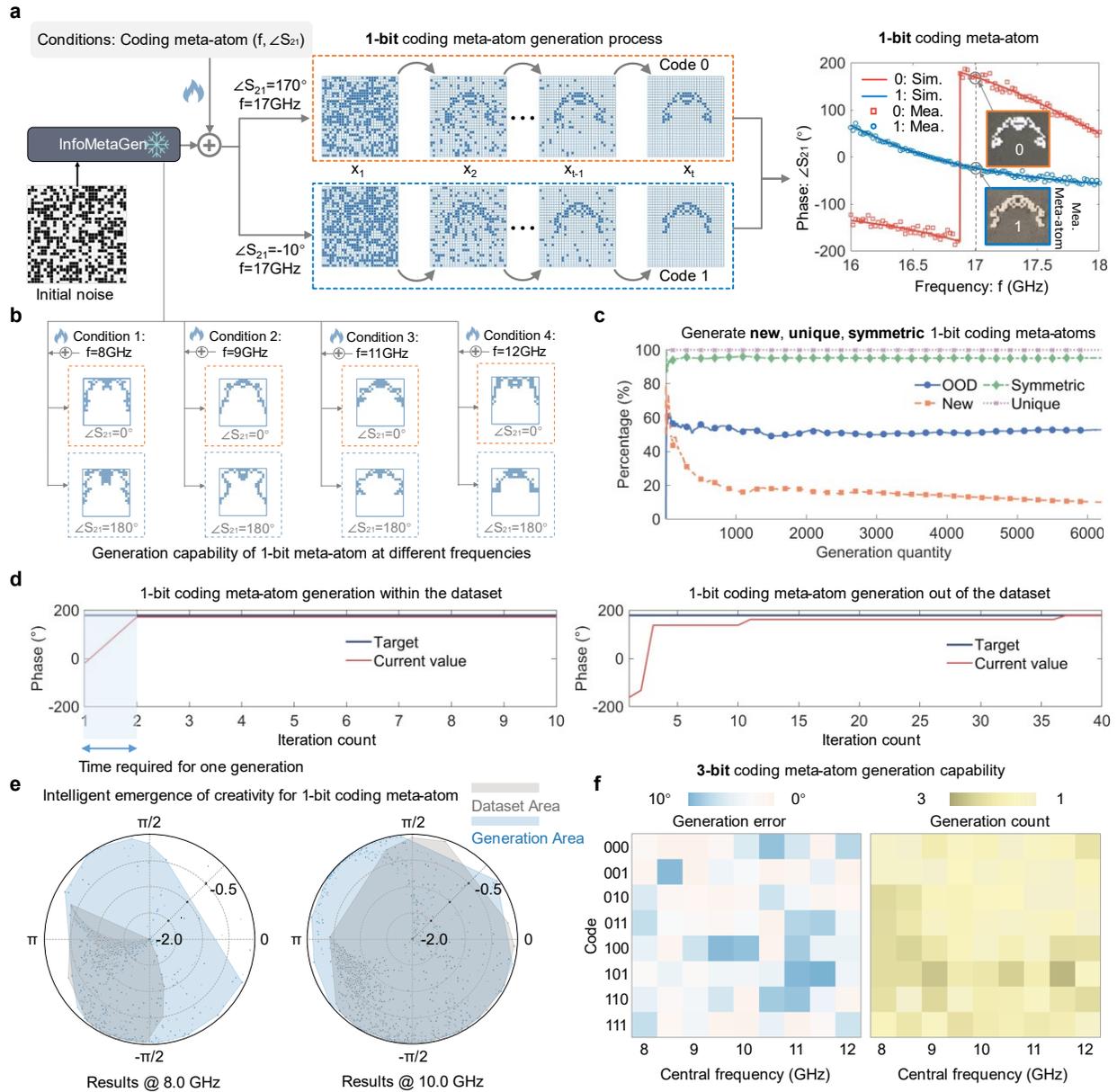

**Fig. 2 | Generation of unique, new, and symmetric meta-atoms. a,** Two 1-bit coding meta-atoms with 180° phase difference is generated by InfoMetaGen from randomly initialized noise, and their performance is validated by real measurements. **b,** Broadband generation of 1-bit meta-atoms by InfoMetaGen, illustrating its generative capabilities across different frequencies. **c,** Percentage of the generated structures classified as unique, symmetric and diversity. **d,** Time required to generate the in-dataset versus out-of-dataset structures. **e,** Creative generation capacity of the 1-bit coding meta-atoms, with the gray area representing the dataset region and blue

area indicating the generable region, highlighting the generative potential. **f,** Generation error and required time for producing 3-bit coding meta-atoms based on a small dataset.

Its exceptional extrapolative performance at the target frequencies of 8 GHz and 10 GHz is demonstrated in in **Fig. 2e**, highlighting the InfoMetaGen's extrapolative generation in unseen design spaces, which is a critical criterion to evaluate the generative models. Notably, although the simulated dataset does not include the meta-atoms that satisfy the criteria for 3-bit coding, InfoMetaGen can still generate the required 3-bit coding meta-atoms across different frequencies, as illustrated in **Fig. 2f**, achieving a phase error below 10° and requiring no more than three inference iterations. These results confirm the InfoMetaGen's strong generative capability for on-demand meta-atom designs.

**Generating far-field beam-forming meta-arrays**

Once the library of coding meta-atoms is generated by InfoMetaGen, we are able to arrange these meta-atoms into a functional meta-array, namely metamaterial. Generally, the functionality of metamaterials can be classified into two categories: far-field manipulations and near-field manipulations. Here, we focus on the far-field beam forming. Since the phase, amplitude and frequency responses of meta-atoms have been precisely characterized and quantified into a discrete coding library, any desired far-field radiation pattern can be synthesized by assigning the optimal coding sequence of meta-atoms. However, current far-field beam-forming optimization still faces significant challenges. Most optimization methods for far-field beam forming require the repeated executions of commercial simulation software in high-dimensional mixed discrete-continuous spaces, resulting in substantial computational costs.

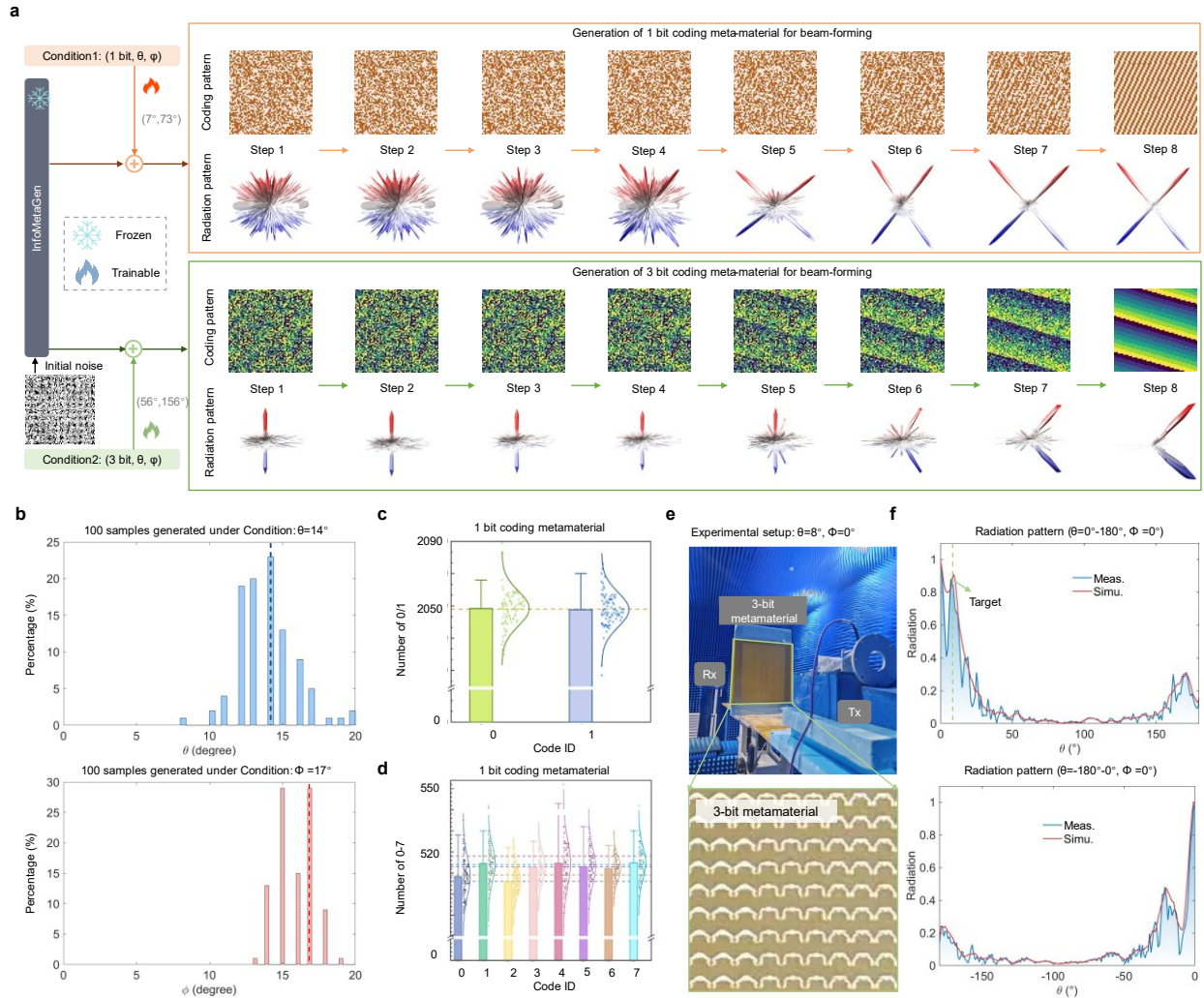

**Fig. 3 | Generation of meta-arrays for far-field beam manipulations. a,** 1-bit (top) and 3-bit (bottom) coding array generations by InfoMetaGen and the corresponding 3D far-field radiation patterns. **b,** Statistical results from 100 independent generations, evaluated against the target angle of $\theta = 14°$ and $\varphi = 17°$. **c, d** The distribution of digit counts for the 1-bit ('0' and '1', **c**) and 3-bit ('0'-'7', **d**) beam steering coding patterns from 100 independent generations under the condition in **b**. Bar plots show the mean digit counts, with the error bars indicating the standard deviation; dashed horizontal lines denote the corresponding simulated counts. **e,** Photographs of the experimental setup and fabricated information metamaterials used for testing the beam steering. **f,** The comparison between simulated and measured results under the condition of $\theta = 8°$ and $\varphi = 0°$.

Now we evaluate the generative capability of InfoMetaGen for the far-field beam steering. Building upon the pretrained diffusion base model, we finely tune a functional adapter that is conditioned on spherical coordinates $(\theta, \varphi)$ and bit type. Then InfoMetaGen generates arbitrary-bit coding patterns on demand, which is evaluated with 1-bit and 3-bit coding metamaterials. **Figure 3a** illustrates the coding pattern generations through the reverse denoising process, which is initialized from random Gaussian noise, along with the corresponding 3D far-field radiation patterns for 1-bit (top) and 3-bit (bottom) coding metamaterials. In the early stages, the random noise is gradually organized into emerging stripe-like motifs. With continued denoising, the patterns are rapidly converged to well-arranged coding patterns, yielding the far-field radiations with distinct beams accurately satisfying the target deflection $(\theta, \varphi)$ for both 1-bit and 3-bit cases. **Figure 3b** gives a statistical analysis of 100 independent generations under identical conditions, showing that the generation errors remain within $3°$, demonstrating the model's high accuracy and robustness in generating the beam-steering meta-arrays. **Figure. 3c and 3d** further examines the statistical distribution of digits in the 1-bit (0 and 1) and 3-bit (0-7) coding patterns from 100 independent generations. Though digit counts vary across generations, they consistently fluctuate around the simulated reference values, indicating that InfoMetaGen can reproduce stable beam steering coding patterns closely matching to the target digital compositions while maintaining the generative diversity. Additionally, we conduct experimental validation, with photographs of the setup and fabricated samples shown in **Fig. 3e**, and all tests are performed in a far-field microwave anechoic chamber. The good agreement between the simulated and measured results further validates the InfoMetaGen's practical efficacy. As presented in **Fig. 3f**, the primary beam deflection occurs at 8 degrees, with a minor peak observed at $172°$. This smaller peak arises primarily because the meta-atoms do not achieve a 100% transmission coefficient, resulting in a residual reflection lobe at $172°$.

**Generating near-field focusing and holographic meta-arrays**

In addition to the far-field beam forming, information metamaterials can also manipulate the near-field distributions. One of the core functionalities is near-field focusing[39], by concentrating the incident plane waves into desired focal spots. This capability is essential for a wide range of applications, such as wireless energy transmission, targeted power delivery to implantable device inside bodies, high-resolution near-field imaging and sensing, and security screening systems. However, unlike the far-field beams that only require phase-gradient control, near-field focusing demands precise controls of both phase and amplitude. The traditional sequential optimization approaches are based on the "unit-equivalent-array" assumption, thus they often converge to local optima in high-dimensional hybrid parameter spaces involving continuous and discrete variables, resulting in several hours optimization for a single design iteration.

Here, we demonstrate the InfoMetaGen's capability to design information metamaterials for near-field focusing. Leveraging the pretrained diffusion base model as a foundation, we finely tune a functional adapter using 3D focal coordinates $(x, y, z)$ and bit type as conditions, enabling InfoMetaGen to generate the desired coding patterns at arbitrary-bit settings, as validated by 1-bit and 3-bit metamaterial designs. **Figure 4a** depicts the coding-pattern generation via reverse denoising from random Gaussian noise, together with the corresponding field distributions of the 1-bit (top) and 3-bit (bottom) metamaterials at specified focal point $(x, y, z)$. Initially, disordered noise pregressively evolves into ring-like structures, accompanied by irregular field distributions without clear focal concentration. As the denoising process goes forward, well-ordered coding patterns are rapidly generated, with the resulting field distributions exhibiting distinct focal spots at the prescribed spatial coordinates.

A statistical analysis of 100 independent generations under identical conditions is illustrated in **Fig. 4b**, showing the focusing positional deviations to be confined within 2 units along the $x$-axis and 1 unit along the $y$-axis, respectively. These results confirm high accuracy and good robustness of InfoMetaGen in generating the near-field focusing meta-arrays. **Figure 4c and 4d** present the statistical distribution of digits in 1-bit and 3-bit coding patterns derived from 100 independent generations. While the digit counts are varied across trials, they consistently fluctuate around the simulated reference values. These findings demonstrate that InfoMetaGen reliably reproduces the desired near-field focusing coding patterns, aligned with the simulated digital composition, while exhibiting generative diversity essential for the inverse design. We conduct experimental validation, with photograph of the experimental setup shown in **Fig. 4e**. As illustrated in **Fig. 4f**, the good agreement between measured and simulated results further corroborates the practical efficacy of InfoMetaGen. Together, these results establish InfoMetaGen as a unified framework for achieving high-performance, stable and diverse metamaterial designs for near-field focusing.

When the focal point is further expanded to arbitrary two-dimensional (2D) or 3D intensity distributions, the near-field focusing is naturally evolved into near-field holography[40-42], which enables the projection of complex electromagnetic patterns. This functionality has transformative applications such as super-resolution medical imaging and electromagnetic-screen augmented reality terminals. However, the near-field holography constitutes a fully functional "universal set", with the design space expanding exponentially, as it requires simultaneous optimizations of all encoding meta-atoms. Conventional global search strategies relying on gradient descent or genetic algorithms are very computationally prohibitive, as each iteration generally demands either numerical simulation or experimental calibration, leading to the optimization processes to be several days or even weeks.

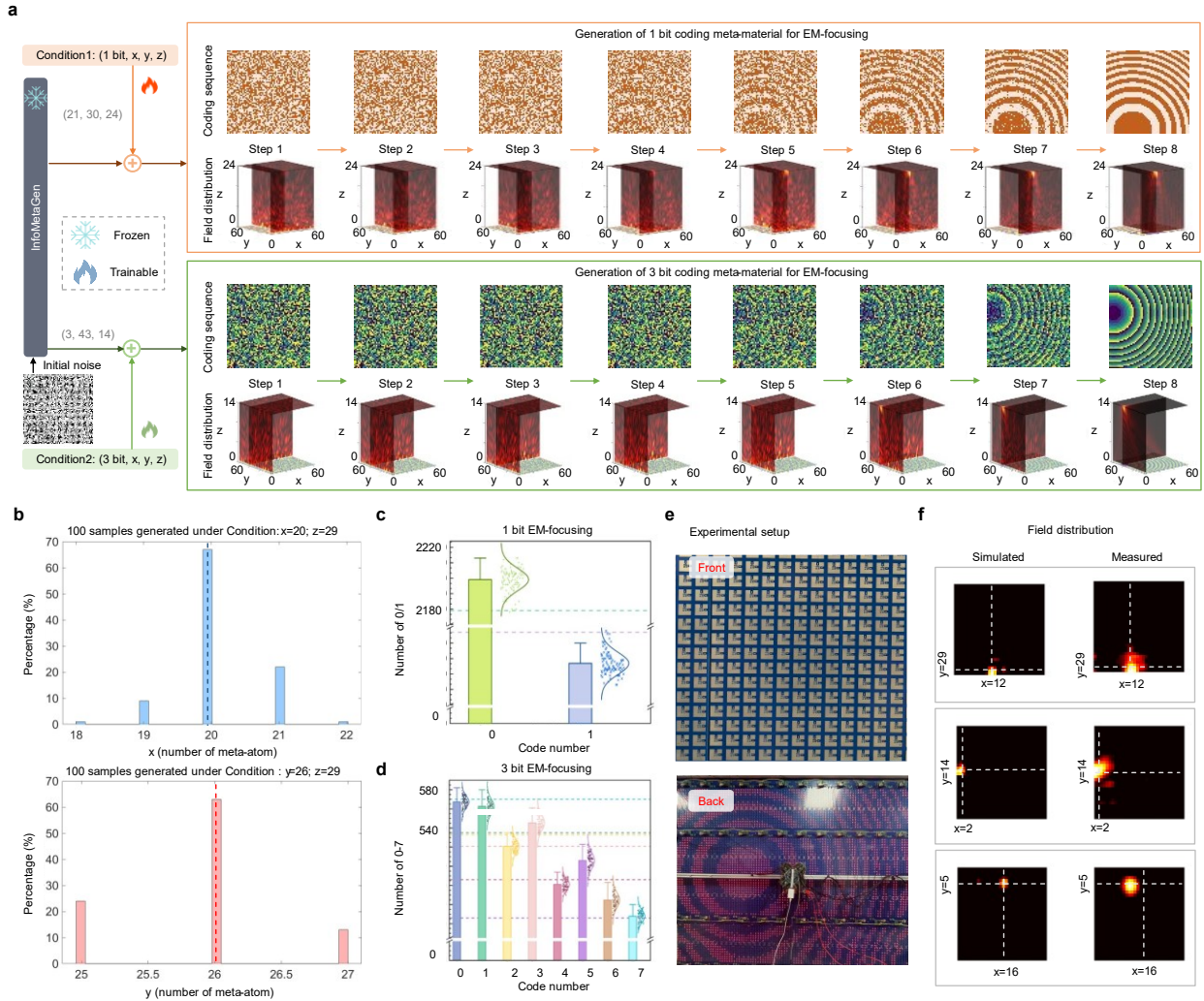

**Fig. 4 | Generation of metamaterials for near-field focusing. a,** Schematic of coding patterns generated by InfoMetaGen and the corresponding 3D near-field distributions for 1-bit (top) and 3-bit (bottom) coding metamaterials at specified focal points in space, represented by the spatial coordinates $(x, y, z)$. **b,** Statistical results from 100 independent generations, evaluated against the focal point of $x = 20$, $y = 26$, and $z = 29$. **c,d,** Distributions of digit counts for the 1-bit ('0' and '1', **c**) and 3-bit ('0'-'7', **d**) near-field focusing coding patterns from 100 independent generations under the condition in b. **e,** Photographs of the experimental setup and fabricated coding metamaterials for near-field focusing testing. **f,** Comparison between simulated and measured results under three different conditions.

Now we evaluate the generative applicability of InfoMetaGen in handling the complicated functionality. We finely tune a functional adapter conditioned on the imaging target and bit type and build up the pretrained diffusion base model. Once finely tuned, InfoMetaGen generates the coding patterns on demand at arbitrary-bit setting, as demonstrated by 1-bit and 3-bit holographic metamaterials. Notably, the imaging distance is not used as a conditional parameter, and the fine-tuning is conditioned on the imaging target. **Figure 5a** illustrates the coding pattern generations via reverse denoising from initial random Gaussian noises, along with the resulting holographic images formed on a cross section of 10 wavelengths in front of the metamaterial, in which the 1-bit (top) and 3-bit (bottom) coding metamaterials are targeting respectively the imaging letters A and C. We note that the initial random noises are progressively organized from disordered motifs into well-defined coding patterns, with the corresponding holographic images displaying clear results. **Figure 5b** depicts the peak signal-to-noise ratio (PSNR) and structural similarity index measure (SSIM) statistics of the resulting holographic images from 100 independent generations under an identical condition (top), and from 20 different conditions (each computed by averaging 100 trials), with higher values indicating greater image similarity. In the statistical plots, PSNR remains above 40dB across all 100 generations, while SSIM consistently exceeds 0.99, showing high accuracy, reliability, and robustness of InfoMetaGen in generating the holographic meta-arrays.

In **Fig. 5c and 5d**, we analyze the digital composition of the generated coding patterns, with 1-bit consisting of digits 0 and 1 and 3-bit consisting of digits 0-7, derived from 100 independent generations under identical conditions. For both 1-bit and 3-bit coding patterns, the digit counts are varied across generations. For the 1-bit trials, the digit counts exhibit small fluctuations, with the mean closely approaching the simulated reference values, indicating good agreement with the target digital composition. In contrast, the 3-bit coding patterns exhibit broader distributions, as revealed by larger standard deviation error bars. This variability does not indicate a failure but rather confirms that InfoMetaGen is capable of generating diverse and novel coding patterns that satisfy the same condition. The experimental apparatus for the near-field holographic imaging is shown in **Fig. 5e**, which is used to validate the 1-bit reflective meta-array and 3-bit transmissive meta-array. **Figure 5f** compares the numerical and measured results, and the high consistency confirms the efficacy and practicality of InfoMetaGen in generating the near-field holographic meta-arrays.

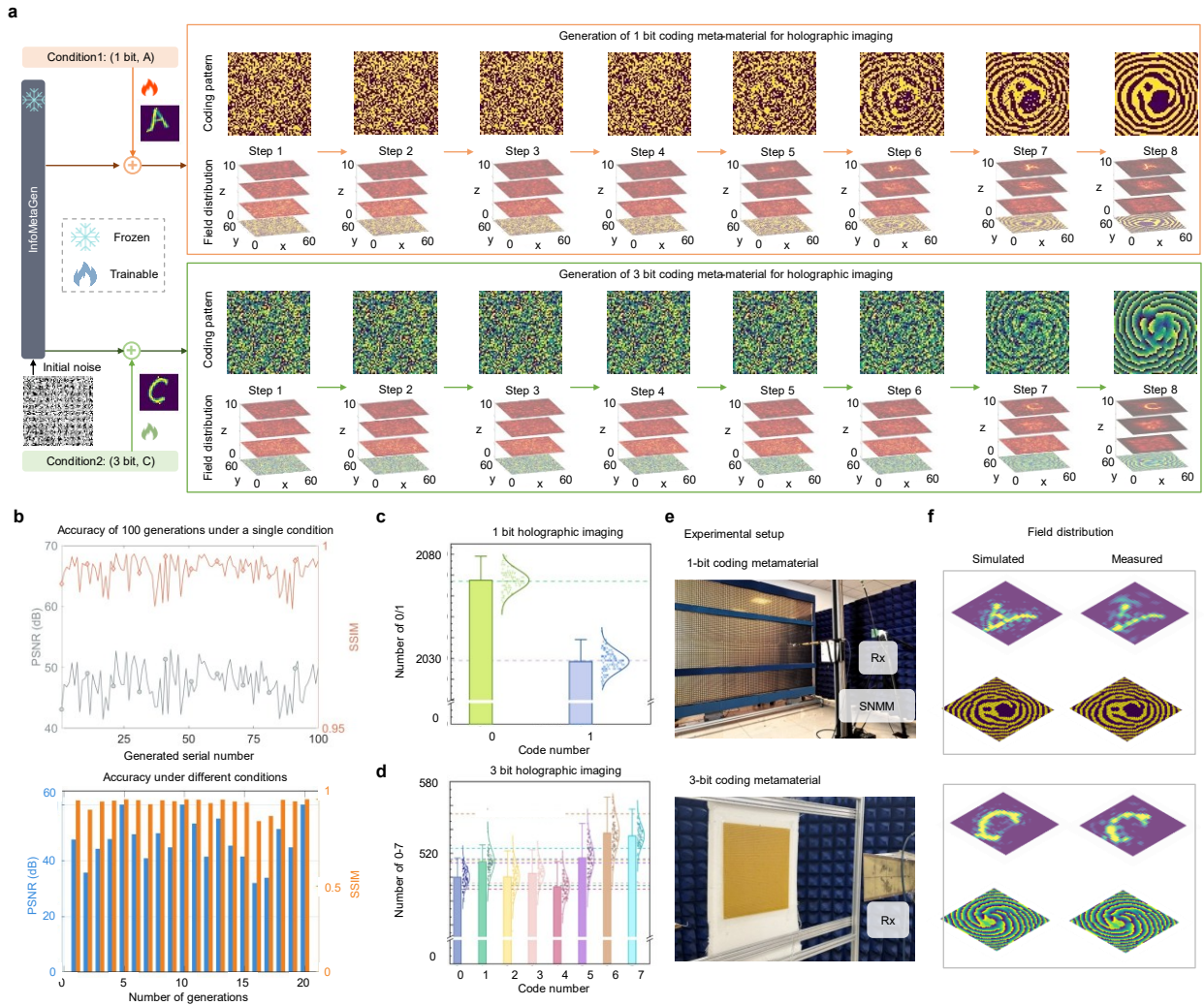

**Fig. 5 | Generation of meta-arrays for near-field holographic imaging. a,** Schematic illustration of coding pattern generations by InfoMetaGen, along with the corresponding holographic images formed on an imaging plane located in front of the meta-array. **b,** PSNR and SSIM statistics from 100 independent generations under an identical condition (top), and from 20 different conditions with each computed by averaging 100 trials (bottom). **c,d,** Distributions of digit counts for the 1-bit ('0' and '1', **c**) and 3-bit ('0'-'7', **d**) holography coding patterns from 100 independent generations. **e,** Photographs of experimental setup and fabricated samples for validating the 1-bit and 3-bit holographic coding meta-arrays. **f,** Comparison between numerical and measured results of the target imaging letter 'A' for the 1-bit meta-array and 'C' for the 3-bit meta-array.

## Discussion

We proposed InfoMetaGen, a unified generative model that enables intelligent inverse designs of information metamaterials across multiple scales, spanning from individual meta-atoms to meta-arrays. Unlike the prior generative models developed for microscopic natural materials (e.g., AlphaFold for protein structure prediction and MatterGen for inorganic crystal design), the proposed InfoMetaGen addressed the inverse design challenges in macroscopic metamaterials, where the current AI-based design models are mostly limited to single-functionality applications. Based on a single foundation model, InfoMetaGen achieved multi-task functionalities, including novel meta-atom generations, beam-steering meta-array generations, electromagnetic focusing meta-array generations, and holographic-imaging meta-array generations. This versatility is achieved not by training separate task-specific models, but by finely tuning lightweight adapters for each specific function, demonstrating InfoMetaGen's remarkable efficiency and adaptability. For the meta-atom designs, InfoMetaGen generated novel, symmetric, and unique meta-atoms, exhibiting exceptionally diverse creativity. More importantly, it produced high-performance meta-atoms over a broad bandwidth, surpassing the constraints of the training dataset. This intelligent generative capability facilitates efficient discovery of high-performance and free-form meta-atoms. For the meta-array designs, our InfoMetaGen framework also demonstrates strong universality and high accuracy for both far-field and near-field electromagnetic manipulations, which is a much more powerful capability since the meta-array involves inhomogeneous space distributions of meta-atoms.

Although InfoMetaGen established a unprecedented unified paradigm for inverse designs of the information metamaterials across scales and functionalities, further efforts are necessary to broaden the diversity of manipulable physics phenomena, and enhance the accuracy and efficiency of the generative process. We believe that this work opens new avenues for the inverse designs of macroscopic metamaterials, bridging the materials science, information science, and artificial intelligence to accelerate innovations across these interdisciplinary fields.

## Methods

**Bit representations of discrete coding digits**

To address the discrete nature of coding patterns of the information metamaterial, coding digits are expressed as bit representations, which can then be modeled within the standard continuous diffusion framework. Specifically, each discrete coding digit in a coding pattern is firstly represented as binary bits, and then cast into real-valued representations termed analog bits. Next, a standard continuous diffusion model is trained directly on these analog bits. Formally, let a discrete coding digit $d \in \{0,1,\cdots,K-1\}$. We represent it using binary expansion of length $n = [log_2 K]$, where $n$ corresponds to the bit number of metamaterial coding pattern. Such that, $d \in \mapsto b = (b_1, b_2, \cdots, b_n)$, $b_i \in \{0,1\}$. Each binary bit $b_i$ is further mapped into analog bit $a_i$ through the transformation : $a_i = 2b_i - 1$, $a_i \in \{-1, +1\}$. Thus, a discrete coding digit $d$ is represented as an analog vector $d \in \mapsto a = (a_1, a_2, \cdots, a_n)$, $a \in \{-1, +1\}$, enabling direct modeling within a continuous diffusion framework. During sampling, InfoMetaGen generates analog bits $\tilde{a}$, which are quantized by thresholding at zero $\hat{b}_i = \mathbf{1}(\tilde{a}_i > 0)$, where $\mathbf{1}(\cdot)$ is the indicator function. The recovered binary vector $\hat{b}$ is then mapped back to the corresponding discrete coding digit $\hat{d}$.

**Pretraining of diffusion base model**

In the pretraining phase, our primary goal is to train an unconditional functionality-agnostic diffusion base model that captures structural distribution of meta-array coding patterns, e.g., their stable structure arrangements and digital compositions. To achieve this, we collect large-scale simulated coding patterns from diverse functionalities spanning from meta-atoms to meta-arrays as the training data. Let a bit-represented coding pattern be denoted as $e \in \mathbb{R}^{H \times W \times n}$, where $H$ and $W$ are the numbers of rows and columns, respectively, and $n$ denotes the maxmium bit number. In the forward diffusion process, pure Gaussian noise $\epsilon \sim \mathcal{N}(0, I)$ is incrementally added to $e$ over $T$ time steps. A U-Net denoising network is then trained to predict the noise at each step, with the optimization objective defined as: $\mathcal{L}_{DM} = \mathbb{E}_{e_0, \epsilon, t}[\|\epsilon - \epsilon_\theta(e_t, t)\|_2^2]$, where $e_t$ is the noisy coding pattern at time step $t$, $\epsilon_\theta(e_t, t)$ denotes the noise predicted by the denoising network parameterized by $\theta$.

**Finely-tuned functional adapter**

After pretraining, the diffusion base model encapsulates rich and diverse prior knowledge across various design spaces, spanning from individual meta-atoms to meta-arrays. Building upon this expressive and robust foundation, we finely tune a lightweight functional adapter to enable task-specific inverse design of meta-array. This eliminates the need to retrain the entire diffusion model for each new task, significantly improving the computational efficiency. In practice, the parameters of the pretrained diffusion base model are frozen, and only a functional adapter is finely tuned to steer the generation process toward the desired functionalities under specific conditions and bit types. The fine-tuning objective follows the same form as pretraining, except that it incorporates the conditional information $c$, as represented below: $\mathcal{L}_{DM} = \mathbb{E}_{e_0,\epsilon,t}[\|\epsilon - \epsilon_\theta(e_t, t, c)\|_2^2]$.

**Numerical simulation of generated metamaterials**

The numerical simulations are primarily focused on the applications of InfoMetaGen in both meta-atoms and meta-arrays. To investigate the meta-atoms, we uesd commercial software CST Microwave Studio for simulations. Due to the periodic nature of meta-atoms, we implement unit cell boundary conditions on all sides, and apply open add space conditions in the z-axis direction to simulate an infinite periodic repetition environment. For the array-level numerical simulations, open add space conditions are similarly applied in the x, y, and z directions for both far-field beam forming and near-field functionalities. In the far-field beam forming, far-field monitors are deployed; for the near-field functionalities, the distance is extended along the positive axis, and electric field monitors are configured.

**Experimental verification of generated metamaterials**

All experimental samples were fabricated using PCB technology, with the patterns composed of 0.018 mm thick copper foil. The dielectric material employed is FR4 (dielectric constant of 4.3, loss tangent of 0.025). In the S-parameter testing experiments for the meta-atoms, the samples consist of a 32 × 32 arrangement of identical meta-atoms. In the lens platform, the upper and lower lenses are connected to a vector network analyzer (model: Agilent N5230C) via two coaxial lines, using the time-domain gating techniques for measurements. In the far-field beam control experiments, the setup is conducted in a far-field anechoic chamber, with the meta-array positioned at the center of a rotating platform. The transmitting horn is located at one end, while the receiving horn is positioned at the opposite end of the chamber. The horn used is model XB-CDL1-18S, with a voltage standing wave ratio (VSWR) of ⩽ 2.5 and SER of 16 092 802#. For the near-field focusing and holographic imaging experiments, the electric field data are collected using a near-field scanning platform, which consists of a two-dimensional scanning frame, a personal computer for control, a vector network analyzer, a receiving probe, a transmitting horn, and two coaxial lines.

## Data Availability

The data that support the findings of this study are available from the corresponding author upon request.

## Code Availability

The code that supports the findings of this study are available from the corresponding author upon reasonable request.

## Acknowledgement

The work was supported by the National Natural Science Foundation of China (Nos. 62288101 and 62301149), National Key Research and Development Program of China (No. 2023YFB3813100), Jiangsu Planned Projects for Postdoctoral Research Fund (No. 2023ZB318), Special Fund for Key Basic Research in Jiangsu Province (Nos. BK20243015, BK20230820).

## Author contributions

J. H, L. C, X. Z, J. Y, and T. C. conceived the study, implemented the methods, performed computational experiments and wrote the paper. J. H, X. Z, J. W, J. Z, and J. Y. helped with the development of functional adapters, model training, and the video illustration. Z. C, L. C, J. S, J. Y, L. L. and T. C. assisted with the experimental verification and preparation of the supplementary material. C. W, L. C, J.Y and T. C. contributed to the figure visualizations.

## Competing interests

The authors declare no competing interests.

## Additional information

**Supplementary information** The online version contains supplementary material available at https://doi.org/XXX.

**Correspondence** and requests for materials should be addressed to J. W. You, J. N. Zhang and T. J. Cui.